\documentclass[sigconf]{acmart}

\usepackage{algorithmic}
\usepackage[ruled,linesnumbered,vlined]{algorithm2e}
\usepackage{setspace}
\usepackage{listings}
\usepackage{multicol}
\usepackage{multirow}
\usepackage{makecell}
\usepackage{amsthm}
\usepackage{enumitem}
\usepackage{tikz}
\usepackage{xspace}
\usepackage{wrapfig}
\usepackage{graphicx,calc, scalerel}

\renewcommand\footnotetextcopyrightpermission[1]{}

\newcommand{\parabf}[1]{\noindent\textbf{#1}}

\newcommand{\CodeIn}[1]{{\small \texttt{#1}}}
\newcommand{\Comment}[1]{}

\newcommand{\tech}{\textsc{ChatRepair}\xspace} %
\newcommand{\baserepair}{BaseChatGPT\xspace} %

\newcommand{\dfj}{Defects4j\xspace}
\newcommand{\quixbugs}{QuixBugs\xspace}

\newcommand{\codexrepair}{CodexRepair\xspace}
\newcommand{\alpharepair}{AlphaRepair\xspace}
\newcommand{\rewardrepair}{RewardRepair\xspace}
\newcommand{\selfapr}{SelfAPR\xspace}
\newcommand{\recoder}{Recoder\xspace}

\newcommand{\cure}{CURE\xspace}
\newcommand{\coconut}{CoCoNuT\xspace}
\newcommand{\dlfix}{DLFix\xspace}
\newcommand{\sequencer}{SequenceR\xspace}
\newcommand{\tbar}{TBar\xspace}
\newcommand{\prapr}{PraPR\xspace}
\newcommand{\avatar}{AVATAR\xspace}
\newcommand{\simfix}{SimFix\xspace}
\newcommand{\fixminer}{FixMiner\xspace}
\newcommand{\capgen}{CapGen\xspace}
\newcommand{\jaid}{JAID\xspace}
\newcommand{\sketchfix}{SketchFix\xspace}
\newcommand{\nopol}{NOPOL\xspace}
\newcommand{\jgenprog}{jGenProg\xspace}
\newcommand{\jmutrepair}{jMutRepair\xspace}
\newcommand{\jkali}{jKali\xspace}
\newcommand{\genprog}{GenProg\xspace}

\newcommand{\codex}{Codex\xspace} 
\newcommand{\codegen}{\textsc{CodeGen}\xspace}
\newcommand{\chatgpt}{ChatGPT\xspace}
\newcommand{\instructgpt}{InstructGPT\xspace}
\newcommand{\gpt}{GPT\xspace}
\newcommand{\codebert}{CodeBERT\xspace}
\newcommand{\ctfive}{CodeT5\xspace}

\newcommand{\gptneox}{GPT-NeoX\xspace}
\newcommand{\incoder}{\textsc{InCoder}\xspace}

\newcommand{\rlhffull}{reinforcement learning from human feedback\xspace}
\newcommand{\rlhf}{RLHF\xspace}
\newcommand{\llm}{LLM\xspace}
\newcommand{\llmfull}{Large Language Model\xspace} %
\newcommand{\nlp}{NLP\xspace}
\newcommand{\apr}{APR\xspace}
\newcommand{\aprfull}{Automated Program Repair\xspace}
\newcommand{\msp}{MSP\xspace}
\newcommand{\mspfull}{Masked Span Prediction\xspace}
\newcommand{\mlm}{MLM\xspace}
\newcommand{\mlmfull}{Masked Language Modeling\xspace}

\newcommand{\clmfull}{Causal Language Modeling\xspace}
\newcommand{\nmt}{NMT\xspace}
\newcommand{\nmtfull}{Neural Machine Translation\xspace}

\newcommand{\nlpfull}{Natural Language Processing\xspace}

\newcommand{\relevanttestcode}{relevant test code\xspace}

\newcommand*\person{%
\scalerel*{\includegraphics{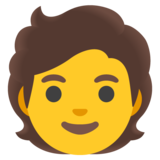}}{\strut}\xspace%
}

\newcommand*\robot{%
\scalerel*{\includegraphics{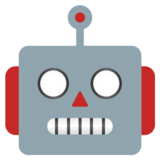}}{\strut}\xspace%
}

\newcommand*\gear{%
\scalerel*{\includegraphics{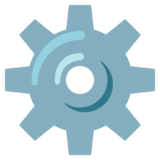}}{\strut}\xspace%
}

\newcommand{\distance}{5pt}
\begin{document}

\title{Keep the Conversation Going: \\ Fixing 162 out of 337 bugs for \$0.42 each using ChatGPT}

\author{Chunqiu Steven Xia}
    \affiliation{\institution{University of Illinois Urbana-Champaign}\country{}}
    \email{chunqiu2@illinois.edu}
\author{Lingming Zhang}
    \affiliation{\institution{University of Illinois Urbana-Champaign}\country{}}
    \email{lingming@illinois.edu}

\begin{abstract}
\aprfull (\apr) aims to automatically generate patches for buggy programs. Traditional \apr techniques suffer from a lack of patch variety as they rely heavily on handcrafted or mined bug fixing patterns and cannot easily generalize to other bug/fix types. To address this limitation, recent \apr work has been focused on leveraging modern \llmfull{s} (\llm{s}) to directly generate patches for \apr. Such \llm-based \apr tools work by first constructing an input prompt built using the original buggy code and then querying the \llm to either fill-in (cloze-style \apr) the correct code at the bug location or to produce a completely new code snippet as the patch. While the \llm-based \apr tools are able to achieve state-of-the-art results, it still follows the classic Generate and Validate (G\&V) repair paradigm of first generating lots of patches by sampling from the same initial prompt and then validating each one afterwards. This not only leads to many repeated patches that are incorrect but also miss the crucial and yet previously ignored information in test failures as well as in plausible patches.

To address these aforementioned limitations, we propose \tech, the first \emph{fully automated} \emph{conversation-driven} \apr approach that interleaves patch generation with instant feedback to perform \apr in a conversational style. \tech \emph{first feeds the \llm{} with relevant test failure information to start with, and then learns from both failures and successes of earlier patching attempts of the same bug for more powerful \apr}. For earlier patches that failed to pass all tests, we combine the incorrect patches with their corresponding relevant test failure information to construct a new prompt for the \llm{} to generate the next patch. In this way, we can avoid making the same mistakes. For earlier patches that passed all the tests (i.e., plausible patches), we further ask the \llm to generate alternative variations of the original plausible patches. In this way, we can further build on and learn from earlier successes to generate more plausible patches to increase the chance of having correct patches. While our approach is general, we implement \tech using state-of-the-art dialogue-based \llm{} -- \chatgpt. Our evaluation on the widely studied \dfj dataset shows that \tech is able to achieve the new state-of-the-art in repair performance, achieving 114 and 48 correct fixes on \dfj 1.2 and 2.0 respectively. By calculating the cost of accessing \chatgpt, we can fix 162 out of 337 bugs for \$0.42 each!

\end{abstract}
\maketitle

\section{Introduction}

\aprfull (\apr)~\cite{gazzola2019aprsurvey, goues2019automated} is a promising approach to automatically generate patches for bugs in software. Traditional \apr tools often use the Generate and Validate (G\&V)~\cite{long2016analysis} paradigm to first generate a large set of candidate patches and then validate each one against the original test suite to discover a set of \emph{plausible} patches (which pass all the tests). These plausible patches are then given to the developers to find a \emph{correct} patch that correctly fixes the underlying bug. Traditional \apr techniques can be categorized into template-based~\cite{ghanbari2019prapr, hua2018sketchfix, martinez2016astor, liu2019tbar, liu2019avatar}, heuristic-based~\cite{legoues2012genprog, le2016hdrepair, wen2018capgen} and constraint-based~\cite{mechtaev2016angelix, le2017s3, demacro2014nopol, long2015spr} \apr tools. Among these traditional techniques, template-based \apr tools, using handcrafted or mined repair templates to match and fix buggy code patterns, have been regarded as the state-of-the-art~\cite{ghanbari2019prapr, liu2019tbar, benton2020effectiveness}. However, template-based tools suffer from lack of patch variety as they cannot easily generalize to bugs and patterns outside of the list of pre-defined templates.

To address the limitations of traditional \apr techniques, researchers have proposed learning-based \apr approaches that leverage advances in Deep Learning. Learning-based approaches are mainly based on either \nmtfull (\nmt) or \llmfull (\llm). \nmt-based \apr tools~\cite{ye2022selfapr, ye2022rewardrepair, zhu2021recoder, jiang2021cure, lutellier2020coconut, li2020dlfix, chen2018sequencer} view repair as a translation task to turn buggy code into correct code by training a \nmt model~\cite{sutskever2014mt} using a dataset of historical bug fixes. However, such \nmt-based \apr tools rely heavily on its training data, obtained by scraping open-source repositories for bug fixing commits. This means that not only can the training dataset be noisy~\cite{jiang2021extracting} (i.e. containing irrelevant commits/changes) but also that these \nmt-based approaches may not generalize to bug fix types not seen in their limited training data.

\begin{figure}[t]
    \captionsetup{justification=centering}
    \centering
    \includegraphics[keepaspectratio=true,width=0.8\linewidth]{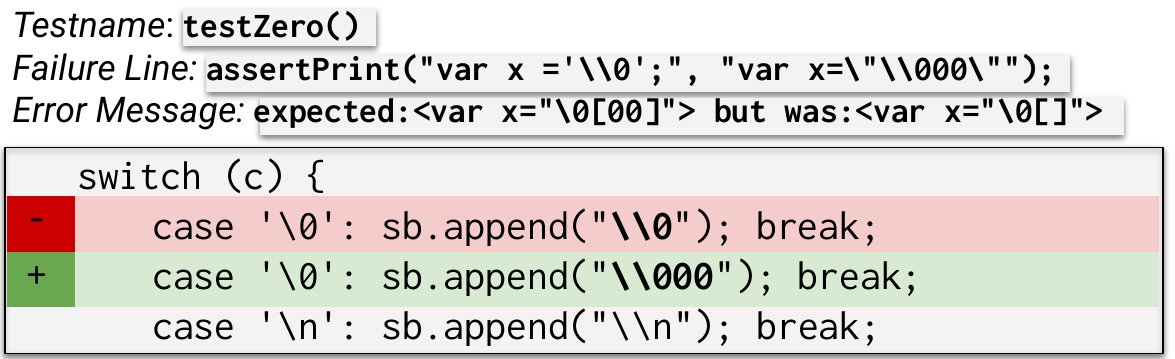}
    \caption{Example bug fix with original testcase information}
    \label{fig:example_fix}
\end{figure}

More recently, researchers have started to directly leverage advanced \llm{s} for \apr~\cite{xia2022alpharepair, xia2023repairstudy, kolak2022patch, prenner2021codexws}. Modern \llm{s} are trained on billions of open-source code snippets, demonstrating impressive performance on many code-related tasks~\cite{brown2020language, codex, fried2022incoder, xu2022systematic}, and can learn to directly generate code given the surrounding context (due to code naturalness~\cite{hindle2012softwarenatural, ray2016natural}). \alpharepair~\cite{xia2022alpharepair} proposes the first cloze-style (or infilling-style) \apr approach, where the buggy code is removed and a \llm directly predicts correct code given the prefix and suffix context. Recent work has also applied \llm-based \apr to autocomplete a single correct line~\cite{kolak2022patch} or to generate a complete fixed function~\cite{prenner2021codexws}. A more extensive study~\cite{xia2023repairstudy} has investigated applying larger \llm{s} and different \llm architectures (i.e. generative and infilling) for \apr, and demonstrates that \llm-based \apr tools can achieve the new state-of-the-art performance on many \apr tasks. Meanwhile, the pipeline for existing \llm-based \apr still has the following limitations:

\emph{1) Missing test failure information.} Current \llm-based tools do not consider the rich information within the original bug-exposing tests. Such information can not only help \llm{s} understand the \emph{meaning} of the source code under tests but can help and hint with concrete code snippets. Figure~\ref{fig:example_fix} shows an example bug fix along with the original test failure information. We see that the fix is to swap the appending string to \CodeIn{"\textbackslash\textbackslash000"}. This can be an extremely difficult fix for \llm-based approaches since this unique string is not a commonly used string seen during pre-training and also there are no other examples of triple strings (\CodeIn{"\textbackslash\textbackslash{}XXX}") within the current function context. However, from the failure line within the test and the corresponding error message, we see that the test expects the output to contain the triple zeros and even contains a code snippet (\CodeIn{"\textbackslash\textbackslash000"}) which is directly used in the patch! \llm{s} have shown powerful in processing/exploiting such unstructured/complex information like test failure logs. By failing to consider them, \llm-based tools may waste a lot of time generating irrelevant patches.

\emph{2) Repeated sampling.} Current \llm-based approaches first construct an input prompt using the original buggy code and either ask the \llm to fill-in the correct code (i.e. cloze-style \apr) or generate a completely new fixed function~\cite{xia2023repairstudy, prenner2021codexws}. Using the initial prompt, \llm-based techniques will sample the \llm multiple times to generate many patches, akin to the traditional G\&V paradigm of program repair. However, since each sample is identically independent, the \llm does not know any previously generated patches. As such, \llm-based tools may generate many repeated or similar patches that were already determined to be incorrect, wasting dollar cost in API access or time in GPU execution. Furthermore, this repeated sampling procedure is also drastically different from how human developers fix bugs, where we iterative build on top of the knowledge and tries from previous failed attempts to come up with the next possible patch.

\emph{3) Ignorance of valuable plausible patches.} In addition to failing to use past incorrect patches, current \llm-based \apr tools also cannot effectively exploit the plausible patches generated earlier. Plausible patches have been shown to be valuable since they often share similar locations with the actual correct patches~\cite{ghanbari2019prapr, lou2020profl}. Moreover, we further hypothesize that plausible patches may also include key code ingredients to pass all tests, and may also help \llm{s} better learn how to pass all tests to generate more plausible patches (thus increasing the chance of generating correct patches). By ignoring such valuable plausible patch information and starting from scratch after generating plausible patches, existing \llm-based \apr may miss opportunities to correctly fix more bugs.

\parabf{Our Work.} We present \tech{} -- a fully automated \emph{conversation-driven} \apr approach that interleaves patch generation with instant feedback to perform patch generation in a conversational style. While our idea is general, to build \tech, we use the recently developed, current state-of-the-art dialogue-based \llm{} -- \chatgpt~\cite{chatgpt}\footnote{While repair uses \chatgpt, no part of this paper is written by \chatgpt.}, which is not only trained on billions of code snippets, but also is designed to be used in a conversational manner to better understand instructions. 
\tech first extracts relevant test failure information to serve as the initial prompt to provide \chatgpt more contextual information for \apr. Moreover, \tech further learns from both failures and successes of earlier patching attempts of the same bug for more powerful \apr. For earlier patches that failed to pass all tests, we combine the incorrect patches with their corresponding test failure information to construct a new prompt for the \llm to generate the next patch. In this way, we can avoid making the same mistakes. For earlier patches that passed all the tests (i.e., plausible patches), we further ask the \llm to generate alternative variations of the original plausible patches. In this way, we can further build on and learn from earlier successes to generate more plausible patches to increase the chance of having correct patches. 
As our approach uses the \chatgpt model, we also compute the dollar cost of \chatgpt API queries used to fix a bug. Surprisingly, we found that \emph{by using \tech, we can fix 162 out of 337 bugs for \$0.42 each.\footnote{This is a reference to a prior classic study done for \apr~\cite{le2012systematic} not using \chatgpt.}}

This paper makes the following contributions:
\begin{itemize}[noitemsep, leftmargin=*, topsep=0pt]
    \item \textbf{Dimension.} We open a new dimension of conversation-driven paradigm for fully automated program repair. Our work demonstrates for the first time that we can effectively leverage previously ignored test failure information, as well as earlier patch attempts in a conversational manner to prompt \llm{s} to generate more correct patches. Moreover, we show the promising future of leveraging dialogue-based \llm{s} for \apr in general. 
    \item \textbf{Technique.} We develop \tech, a \emph{fully automated} conversation-driven \apr tool using the very recent \chatgpt model. More specifically, we automatically extract concise and relevant information about the initial test failures as well as earlier patch attempts to prompt \chatgpt for effective \apr.%
    \item \textbf{Evaluation.} We evaluate \tech against current state-of-the-art learning-based and traditional \apr tools on the widely studied \dfj 1.2, 2.0~\cite{just2014dfj} and \quixbugs~\cite{lin2017quixbug} dataset. \tech obtains the new state-of-the-art repair result of 114 and 48 correct bug fixes (15 and 17 more than prior best baseline) on \dfj 1.2 and 2.0 respectively. Additionally, we conduct an extensive ablation study to demonstrate the improvement gained from both utilizing rich semantic test failure information and the conversational paradigm of \tech for repair.
    
\end{itemize}

\section{Background \& Related Work}

\subsection{\llmfull}

\llmfull{s} (\llm{s})~\cite{brown2020language} have seen meteoric rise in both performance and corresponding adoptions due to recent advances in \nlpfull (\nlp) that enable scaling \llm size to billions of parameters and using billions of training samples. As \llm{s} are trained to be general and can capture knowledge from various different domains, \llm{s} are either \emph{fine-tuned}~\cite{radford2018improving} or \emph{prompted}~\cite{liu2023pre} in order to be used for a downstream task. Fine-tuning involves updating the model parameters with a specific training dataset to target a particular downstream task. However, fine-tuning is not only expensive as it requires additional model training, but may also be infeasible in cases where sufficient training datasets are unavailable. Prompting on the other hand directly uses \llm{s} without any training by providing natural language descriptions of the downstream task (e.g., produce a docstring) and optionally a few demonstration of the task being solved as input to the \llm. 

\llm{s} are built on the transformer architecture~\cite{vaswani2017attention} and can be classified based on the component(s) used. Decoder-only models (e.g., \codex~\cite{codex} and \codegen~\cite{codegen}) are the popular \gpt-based models trained using \clmfull objective by training to predict the probability of the next token given all previous left only context. Encoder-only (e.g., \codebert~\cite{feng2020codebert}) and Encoder-Decoder (e.g., \ctfive~\cite{wang2021codet5}) models are trained using \mlmfull (\mlm) or \mspfull (\msp) objective, respectively, where a small portion (e.g., 15\%) of the tokens are replaced with either masked tokens or masked span tokens and the \llm{s} are trained to recover the masked out tokens based on bi-directional context. 

In addition to these traditional \llm{s}, more recently, researchers have proposed \llm{s} trained using reinforcement learning which \emph{aligns} better with human preference~\cite{ouyang2022instructgpt, chatgpt, ziegler2019rlhf}. Examples include \instructgpt~\cite{ouyang2022instructgpt} and \chatgpt~\cite{chatgpt} which are first initialized from a pre-trained model on autoregressive generation and then fine-tuned using \rlhffull (\rlhf)~\cite{ziegler2019rlhf}. \rlhf first fine-tunes the base model using a small dataset of prompts (input) and desired output (human-written). Then a separate reward model is trained on a larger set of prompts by sampling multiple outputs from the fine-tuned \llm and using a human labeler to rank each individual output. Finally, reinforcement learning (e.g., Proximal Policy Optimization~\cite{schulman2017ppog}) is applied to calculate the reward of the output generated based on the reward model and correspondingly update the \llm weights. The resulting \llm through fine-tuning using human preference has shown better understand complex input prompts and follow instructions to perform various tasks~\cite{ouyang2022instructgpt, chatgpt, bang2023multitask}. Specifically, \chatgpt has received lots of attention due to its dialogue/conversation focus by training specifically on conversations and its ability to keep track of and reference prior conversations.  

In this work, we continue to build on our in-progress work~\cite{xia2023conversational}: we introduce a more comprehensive approach that includes more robust feedback (with multi-dimensional relevant test information) and aims to learn from both failing and plausible patches; we further fully evaluate the ability of RLHF-based \llm{s} for fixing bugs of real-world systems (e.g., \dfj~\cite{just2014dfj}).
This work demonstrates for the first time that powerful RLHF-based \llm{s} like \chatgpt can be directly applied for fully automated conversation-driven \apr, and can substantially outperform all existing \apr techniques.%

\subsection{\aprfull}

\aprfull (\apr) can help developers by generating patches for a given bug based on its potential fault location(s). Classic \apr techniques can be mainly classified as heuristic-based~\cite{legoues2012genprog, le2016hdrepair, wen2018capgen}, constraint-based~\cite{mechtaev2016angelix, le2017s3, demacro2014nopol, long2015spr} and template-based~\cite{ghanbari2019prapr, hua2018sketchfix, martinez2016astor, liu2019tbar, liu2019avatar} ones. Due to the high number of bugs fixed, template-based \apr tools have been recognized as the state-of-the-art. Meanwhile, such \apr tools leverage human-defined or automatically-mined templates to first match potential buggy code patterns and then apply the corresponding fixes. However, template-based tools can only fix the bugs that fall into their limited set of patterns and therefore cannot generalize to other bug types or fixes. To address this issue, researchers have proposed learning-based \apr techniques by leveraging recent advances Deep Learning. Techniques based on \nmt have been extensively studied in recent years, e.g., \selfapr~\cite{ye2022selfapr}, \rewardrepair~\cite{ye2022rewardrepair}, \recoder~\cite{zhu2021recoder}, \cure~\cite{jiang2021cure} and \coconut~\cite{lutellier2020coconut}. They share the same insight that \apr can be viewed as a \nmt problem which aims to translate buggy code into correct code. In this way, they can learn to generate patches by training \nmt models on a dataset of pairs of buggy and fixed code snippets. While effective, such \nmt-based techniques rely heavily on historical bug-fixing training datasets which are usually obtained from scraping open-source repositories for bug-fixing commits. As such, the training data may include various noises such as irrelevant changes/commits; moreover, in order to reduce such false positives, these datasets focus mainly on small commits which further limit the types of bugs/fixes used for training. As a result, \nmt-based \apr techniques are still limited in the type/number of bugs they can fix. 

To further combat the limitations of \nmt-based tools, researchers have also explored the possibility of directly leveraging \llm{s} to synthesize correct patches. \llm{s}, by pre-training on large amounts of open-source code snippets, can directly synthesize the correct code given the surrounding context without having to translate from the buggy code. \alpharepair~\cite{xia2022alpharepair} is the first tool for cloze-style (or infilling-style) \apr where the buggy line(s) is first replaced with masked tokens and then \llm{s} are used to directly fill-in the correct code based on its context. \alpharepair shows for the first time that \llm-based \apr can outperform the widely studied \nmt-based \apr techniques on real-world systems. Prenner et al.~\cite{prenner2021codexws} and Kolak et al.~\cite{kolak2022patch} also directly used \codex~\cite{codex} to generate a fixed function given the original buggy function or to autocomplete a single line given the prefix code on a small dataset (\quixbugs~\cite{lin2017quixbug}). More recently, Xia et al.~\cite{xia2023repairstudy} conducted an extensive study of \llm-based \apr techniques based on various \llm{s} (e.g., \codex~\cite{codex}, \gptneox~\cite{gpt-neox-20b}, \ctfive~\cite{wang2021codet5}, and \incoder~\cite{fried2022incoder}), and further demonstrated the superiority of \llm-based \apr. 
 Despite the promising results of \llm-based \apr, such existing techniques only focus on the source code under repair without considering the rich semantics in test failure information. Furthermore, prior \llm-based techniques continuously sample from the same initial prompt, failing to utilize knowledge from previous failed or plausible patches. In \tech, we build on our in-progress work~\cite{xia2023conversational} and address limitations of prior \llm-based tools by introducing a conversation-based repair paradigm to incorporate both patch generation history with immediate validation feedback for \apr. 

Prior \apr tools have also leveraged simple patch execution or test information for \apr. \genprog~\cite{legoues2012genprog} is a classic \apr tool that uses an evolutionary algorithm to combine candidate patches that pass more tests together. Constraint-based \apr tools~\cite{long2015spr, demacro2014nopol, nguyen2013semfix, durieux2016dynamoth} have used the underlying testing code to extract and build constraints for patch synthesis. Recently, \rewardrepair~\cite{ye2022rewardrepair} proposes to train a \nmt model with a reward function based on whether a patch in the training set passes compilation or test execution. \selfapr is another \nmt-based \apr tool which encodes the bug-exposing test errors together with the original buggy code as input for \apr. Meanwhile, to our knowledge, \tech is the first work that leverages detailed feedback (e.g., including relevant test code and error messages) for each and every patch validated for conversational \apr. Also, \tech directly leverages \llm{s} for digesting test feedback extraction, which is fully automated/generalizable and can understand deep semantic information.%
Researchers have also attempted to use conversation-like approaches for program repair/synthesis~\cite{austin2021synthesis}. However, they typically require human feedback, while \tech is fully automated. %

\section{Approach}
\label{sec:approach}

\begin{figure*}[t]
    \captionsetup{justification=centering}
    \centering
    \includegraphics[keepaspectratio=true,width=0.9\textwidth]{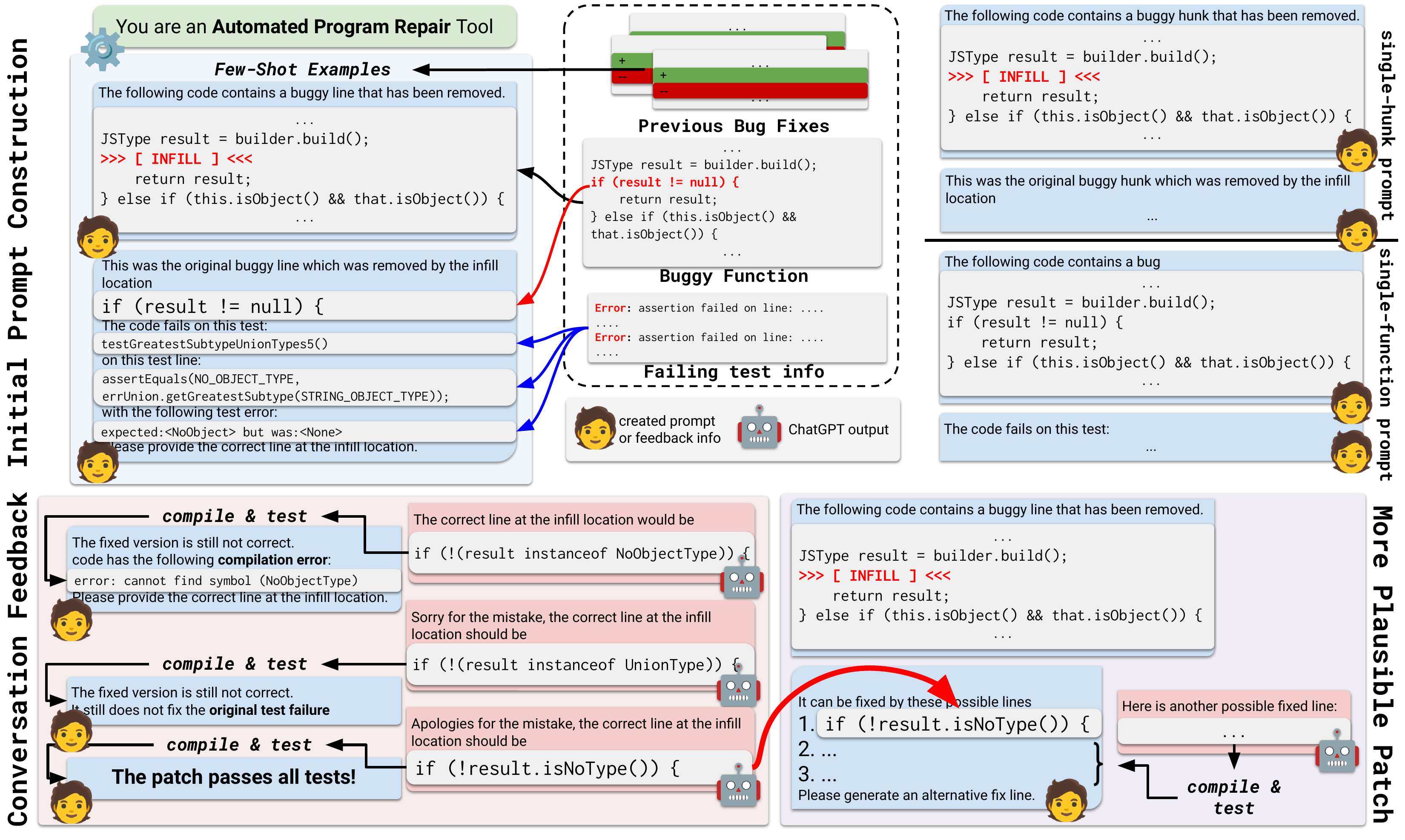}
    \caption{Overview of \tech}
    \label{fig:overview}
\end{figure*}

We propose \tech, a fully automated conversation-driven \apr technique that incorporates multiple dimensions of feedback information to iterative query the model to generate patches. Instead of directly generating patches based on the buggy code as existing \llm-based \apr techniques do, \tech additionally provides valuable test failure information to further assist \llm{s} in patch generation. Moreover, instead of continuously sampling from the same prompt as prior \llm-based \apr techniques do, \tech keeps track of conversation history and further learns from earlier failed and succeeded patching attempts of the same bug via prompting. In this way, \tech can both avoid prior failures and build on earlier successes (e.g., plausible patches) for more effective \apr. As such, \tech maximizes the ability to obtain a genuine correct patch that correctly fixes the underlying bug. While our approach is general and can use different \llm{s} and be applied to a variety of different repair scenarios, in this work, we use the state-of-the-art \chatgpt model~\cite{chatgpt} that is designed specifically for dialogue interaction.

Figure~\ref{fig:overview} shows an overview of \tech using an illustrative repair example. \gear refers to the system message to initialize the model to do a specific task, \person indicates the prompt and feedback \tech provides to the \llm and \robot represents the output response given by \chatgpt. First, \tech initializes \chatgpt with the system message of \textit{"You are an Automated Program Repair tool"} to prepare \chatgpt for the repair task. Then, we construct the initial prompt for \chatgpt which contains the buggy function to be fixed and the relevant test failure information to fix the bug (Section~\ref{sec:initial_prompt}). After querying \chatgpt to generate a potential patch using the initial prompt, we then move onto the conversation stage to first learn from past failures (Section~\ref{sec:conversation}). More specifically, we evaluate the generated patch against the original test suite to see if the patch can pass the previously failed tests. If not, \tech offers immediate feedback by creating a response that includes the relevant failure information (e.g., test failure/compilation error message) and to re-query \chatgpt to generate a new patch while trying to avoid repeating similar failures. This process is repeated until either a plausible patch is produced or the maximum conversation length is reached. After obtaining a plausible patch, \tech attempts to learn from such successes to generate more plausible patches that pass the test suite (Section~\ref{sec:plausible_correct}). \tech prompts \chatgpt with earlier plausible patches to generate more alternative plausible patches. From this process, \tech can obtain multiple plausible patches which can increase the chance of getting the correct patch. We next describe each of the steps in more detail.

\subsection{Initial Input}
\label{sec:initial_prompt}

To begin with, we use the original buggy project and bug to construct our initial prompt \person to \chatgpt to start off the repair process. We follow prior learning-based \apr tools~\cite{xia2022alpharepair, ye2022rewardrepair, jiang2021cure} and focus mainly on line-level repair (specifically infilling or cloze-style \apr as it has been demonstrated to be the state-of-the-art~\cite{xia2022alpharepair}). Meanwhile, \tech is \emph{general} can also be used in a variety of different repair scenarios (see additional prompts in the top-right of Figure~\ref{fig:overview}), which we will evaluate in more detail during later sections.

\newcommand{\model}{\mathcal{C}\xspace}

Figure~\ref{fig:overview} shows an example of an initial prompt. Before we add the target bug to be fixed to the prompt, we first include a few examples of historical bug fixes within the same buggy project. By doing so, we gear the model towards the repair task and allow it to learn the desired output format (i.e. a patch) of the task. After the few-shot examples, we take the original target buggy function to be fixed as the input along with the location of the bug.%
We replace the buggy code within the function with an infill location indicator (\CodeIn{\(\ggg\) [ INFILL ] \(\lll\)}) and refer to this later in the prompt to instruct the model to fill-in the correct code. We then provide the original buggy line which we replace with the infill location indicator to the model since the buggy line can also give useful information as to what a candidate patch should look like. Next, we provide additionally relevant information to help \tech to fix the bug. Our approach uses information derived from the failing test(s) which exposes the original bug. Such bug-exposing tests contain rich semantic information/hints which can help with generating the correct patch to fix the bug~\cite{lou2020profl}.

\newcommand{\prefix}{pre\xspace}
\newcommand{\suffix}{suf\xspace}
\newcommand{\infilltoken}{infill\xspace}
\newcommand{\testinfo}{f_{0}\xspace}
\newcommand{\infillprompt}{I_{fill}\xspace}
\newcommand{\patch}{p\xspace}

\tech uses various information from a failing test, including 1) its name, 2) the relevant code line(s) triggering the test failure, and 3) the error message produced. The name of the failing test can serve as a \emph{short summary} of the function under test. In the Figure~\ref{fig:overview} example, the failing test is \CodeIn{testGreatestSubtypeUnionTypes5()} which tells us that we are testing for a functionality related to the determining greatest subtype from union types. The relevant test code and error message gives concrete information as to why the test failed. In the example, the relevant test code and error message tell the model that we are comparing a \CodeIn{No\_OBJECT\_TYPE}, but the source code function incorrectly returned a \CodeIn{None} object. Such failing test information not only offers the model more explanation in terms of the functionality of the source code but also gives concrete information in terms of expected output and function usage to help the model to generate the correct fix. Note, if there are multiple failing tests, \tech only provides the information from one of them to keep a concise initial prompt. Finally, we end our initial prompt by giving the instruction to model to generate a correct line to replace the buggy code at the infill location. Let \(\model\) be \chatgpt which outputs the probability of generated a sequence, \(\prefix\) and \(\suffix\) as the prefix and suffix of the buggy code with the buggy line removed, \(\infilltoken\) as the infill token replacing the buggy line, \(\testinfo\) as the constructed failure test information and \(\infillprompt\) as the infill instruction prompt. The patch $\patch$ generated can be formalized as the conditional probability: $\model(\patch|\prefix, \infilltoken, \suffix, \testinfo, \infillprompt)$

To our knowledge, \tech is the first work to apply these test failures and error messages in a purely prompting method by combining natural language descriptions of the failure information (e.g., \CodeIn{The code fails on this test: \{failure\_test\}}) as input to the powerful \chatgpt model. Different from prior usage of test execution information for repair~\cite{ye2022selfapr} which relies on custom encodings or handcrafted heuristics, \tech through the use of \chatgpt via prompting is general not only across different programming languages but is also not restricted by the types of test information.

\subsection{Conversational Repair}
\label{sec:conversation}

\newcommand{\altinstruct}{AltInstruct\xspace}

\begin{algorithm}[t!]
\setstretch{0.5}
\small
\caption{Conversational Repair}
\label{alg:dynamic}
\SetKwData{feedback}{feedback}
\SetKwData{patch}{patch}
\SetKwData{ppatch}{pPatches}
\SetKwData{testcase}{oFailure}
\SetKwData{testexecution}{testResult}
\SetKwData{success}{PASS}
\SetKwData{alternateprompt}{\altinstruct}
\SetKwData{initialprompt}{initialPrompt}
\SetKwData{testsuite}{testSuite}
\SetKwData{model}{\chatgpt}
\SetKwData{length}{maxConvLength}
\SetKwData{tries}{maxTries}
\SetKwData{clength}{currentLength}
\SetKwData{ctries}{currentTries}
\SetKwData{input}{input}
\SetKwData{dollarcost}{cost}
\SetKwFunction{validate}{Validate}
\SetKwProg{Fn}{Function}{:}{}
\SetKwFunction{DynamicFeedback}{DynamicFeedback}
\SetKwFunction{ConversationRepair}{ConversationalRepair}
\SetKwFunction{ConstructPrompt}{ConstructPrompt}
\SetKwInOut{Input}{Input}
\SetKwInOut{Output}{Output}
\SetKw{Break}{break}

\Fn{\ConversationRepair}{
    \Input{\initialprompt (initial prompt), \testcase (original failing test info), \testsuite (test suite), \model, \length (max conversation length), \tries (max tries), \alternateprompt (plausible patch prompt)}
    \Output{\ppatch (plausible patches), \dollarcost (total cost)}
    \BlankLine
    \ppatch, \ctries, \dollarcost $\leftarrow$ NONE, 0, \$0\\
    \While{\ctries $<$ \tries and \ppatch is NONE}{ \label{ts:stop}
        \clength $\leftarrow$ 0 \\
        \input $\leftarrow$ \initialprompt \label{ts:initial}\\  
        \While{\clength $<$ \length}{ \label{ts:length}
            \patch, \dollarcost $\leftarrow$ \model(\input) \\
            \testexecution $\leftarrow$ \validate(\patch, \testsuite) \label{ts:validate}\\
            \uIf{\testexecution is \success}{ \label{ts:success}
            \ppatch $\leftarrow$ [ \patch] \\
            \Break
            }
            \uElseIf{\testexecution is \testcase}{ \label{ts:same}
            \feedback $\leftarrow$ "still doesn't fix original failure" \label{ts:same_return}
            }
            \Else{  
            \feedback $\leftarrow$ \ConstructPrompt(\testexecution )\label{ts:different_return}
            }
            \input $\leftarrow$ \{ \input, \patch, \feedback\} \\
            \ctries $\leftarrow$ \ctries$+1$ \\
            \clength $\leftarrow$ \clength$+1$ \\
        }
    }
    \uIf{\ppatch is not NONE}{
        \While{\ctries $<$ \tries}{
            \input $\leftarrow$ \{ \initialprompt, \ppatch, \alternateprompt\} \label{ts:initial_pp}\\
            \patch, \dollarcost $\leftarrow$ \chatgpt(\input) \\
            \testexecution $\leftarrow$ \validate(\patch, \testsuite) \\
            \uIf{\testexecution is \success and \patch not in \ppatch}{ \label{ts:success_pp}
            \ppatch $\leftarrow$ \ppatch | [ \patch] \label{ts:another}\\ 
            }
            \ctries $\leftarrow$ \ctries$+1$ \\
        }
    }
    \algorithmicreturn{ \ppatch, \dollarcost }
}
\end{algorithm}

We first use the initial prompt \person created in Section~\ref{sec:initial_prompt} to query \chatgpt to obtain a model output \robot and extract a candidate patch. Then, we move on to the conversational part of the approach where we interleave patch generation with test validation feedback to prompt future generation in a conversational manner. Each generated patch by the model is followed immediately by a patch validation step to compile and run the patch on the test suite. If the patch failed to pass the test, we construct a detailed feedback information using both the incorrect patch and the failing test as part of the prompt for the next patch generation. Similar to the initial prompt, test failure information can help the model understand the failure reason and provide guidance towards generating the correct fix. In conversation step, we further combine test failure information with previously incorrect patches to not only avoid generating more similarly incorrect patches but also learn from the mistakes of prior generations. We repeat the procedure until a plausible patch which passes the entire test suite is generated.

\newcommand{\previousgenerations}{Q_{<n} = \{(p_1, f_1), ..., (p_{n-1}, f_{n-1})\}\xspace}
\newcommand{\previousgenerationsname}{Q_{<i}\xspace}
\newcommand{\previouspatch}{p\xspace}
\newcommand{\previousfeedback}{f\xspace}
\newcommand{\promptinitial}{I\xspace}

More precisely, we define a \textbf{conversation exchange} as a pair of patch generation and validation feedback of that candidate patch (i.e., \{ \robot, \person\}). Within one repair conversation, the next patch generated by \chatgpt is prompted with the concatenation of the initial prompt with all previous conversation exchanges. For example, the 3rd patch \(\robot_{3}\) is generated with the input being \{ \(\person_{o}\), \(\robot_{1}\), \(\person_{1}\), \(\robot_{2}\), \(\person_{2}\)\}. Let $\model$ be \chatgpt model which outputs the probability of generating a sequence, $\promptinitial$ be the initial prompt, $\previousgenerations$ be the previously generated patch $\previouspatch$ and feedback information $\previousfeedback$ within the same conversation. The next patch generated can be formalized as the conditional probability: $\model(\previouspatch_{i}|\promptinitial, \previousgenerationsname)$

Since \chatgpt (and other \llm{s}) has a limited size context window~\cite{chatgptguide}, meaning it cannot take in arbitrary lengthed inputs, we use \textbf{conversation length} (i.e., the number of exchanges within a single continuous conversation) as another stopping criteria to restart the repair process from the initial prompt once a maximum conversation length is reached. A maximum conversation length of 1 represents the base case of sampling from the initial prompt over and over again and as we increase maximum conversation length, the amount of history (previous patches/feedback) we provide to the model increases.  

Algorithm~\ref{alg:dynamic} details our conversation repair process. Our input includes the initial prompt, original test failure information, test suite, \chatgpt model, plausible patch generation prompt (used in Section~\ref{sec:plausible_correct}), two hyperparameters of maximum conversation length and maximum tries. The final outputs are a list of plausible patches as well as the total cost in \chatgpt API access. Maximum tries is a stopping criteria that stops the repair process once the maximum number of tries (i.e. queries to \chatgpt) has been used to repair a bug (Line~\ref{ts:stop}). Maximum conversation length further limits the maximum amount of prior history used to generate a future patch (Line~\ref{ts:length}). Following the example in Figure~\ref{fig:overview}, we first set the initial input to \chatgpt as the initial prompt (Line~\ref{ts:initial}). \chatgpt first produces the patch which checks if \CodeIn{result} is an instance of \CodeIn{NoObjectType}. This patch may be motivated by the original test failure information within the initial prompt where a similar global constant (\CodeIn{No\_Object\_Type}) is used in the test assertion line. However, this patch is not correct as it contains a compilation error. \tech identifies this by directly attempting to compile and run the test suite and constructing a feedback prompt which indicates that the generated patch has a compilation error (\CodeIn{cannot find symbol (NoObjectType)}). To generate the second candidate patch, \tech concatenates the initial prompt, first generated patch, and validation feedback (indicating the compilation error) together as input to \chatgpt. 

 The second patch indeed fixes the compilation error and checks if \CodeIn{result} is an instance of \CodeIn{UnionType}. \tech employs a \textbf{dynamic} feedback approach as described in Algorithm~\ref{alg:dynamic}. \tech will first compile and run the test suite (Line~\ref{ts:validate}) to observe if the patch can successfully pass the testcases (Line~\ref{ts:success}). In this example, we see that the patch still fails the original bug-exposing test (Line~\ref{ts:same}). Instead of repeating the test error message of the original test, we simply refer back to the initial prompt by saying \CodeIn{It still does not fix the original test failure} (Line~\ref{ts:same_return}). Since we always include the initial prompt in the input, we can provide a concise message to the model to indicate the test failure reason. On the other hand, if the patch can pass the bug-exposing test used in the initial prompt but fails on a different test (either another original failing test or a regression test), we construct the feedback similar to the initial prompt where we include test name, \relevanttestcode, and error message (Line~\ref{ts:different_return}). Note that if there are multiple failing tests, similar to the initial prompt, we only provide feedback for one of them to keep the response succinct. 

 The third patch is generated similar to before where we concatenate the initial prompt with all previous conversation exchanges. We see that in this case the patch which directly calls the member function \CodeIn{isNoType()} is able to successfully pass the test suite. Using the test feedback information such as the error message and the \relevanttestcode, \chatgpt recognizes that this bug deals with a corner case related to none objects or types to generate a plausible patch which fixes the bug.%

\subsection{Plausible Patch Generation}
\label{sec:plausible_correct}

After the previous step, \tech should obtain a plausible patch that can pass the entire test suite. However, a plausible patch may not always be able to correctly fix the underlying bug since the test suite can be incomplete and therefore not cover all possible intended usage of the underlying code~\cite{smith2015cure}. As such, developers have to manually inspect plausible patches to determine correct ones. Both plausible patches and the final correct patches share the similar characteristic: they all can pass the entire test suite. Therefore, instead of starting from scratch (using the buggy code again), \tech directly leverages the existing plausible patch(es) to create more plausible patches. In short, in order to increase the probability that we can generate a correct patch, \tech takes the plausible patches generated previously and asks the model to generate alternative variations and produce additional candidate patches. 

\newcommand{\previousplausible}{PL_{<n} = \{pl_1, ..., pl_{n-1}\}}
\newcommand{\previousplausiblename}{PL_{<i}}
\newcommand{\previousplausiblepatch}{pl}
\newcommand{\altinstruction}{I_{pl}}

Figure~\ref{fig:overview} shows how our plausible patch generation process works. To begin with, we take the initial prompt used (Section~\ref{sec:initial_prompt}) which contains the original buggy code function along with useful test failure information. We then append the prompt with a list of plausible patches generated (Line~\ref{ts:initial_pp} in Algorithm~\ref{alg:dynamic}). In the beginning, this list will only contain the single plausible patch from the previous step, however it grows as we continue to generate additional plausible patches. Next, we indicate in the prompt (\altinstruct in Algorithm~\ref{alg:dynamic}) of the task we want to solve -- \CodeIn{Please generate an alternative fix line.} We then use this prompt as input to \chatgpt and obtain a candidate patch which we will again compile and run the test suite to check if it is indeed another plausible patch (Line~\ref{ts:success_pp}). We continuously query \chatgpt and update our prompt to include new plausible patches generated in order to avoid repeatedly generating the same plausible patch again and also further build on earlier plausible patches (Line~\ref{ts:another}). Again let $\model$ be \chatgpt model which outputs the probability of generating a sequence, $\promptinitial$ be the initial prompt, $\altinstruction$ as the task instruction, $\previousplausible$ be the previous generated plausible patch. The next plausible patch generated can be formalized as the conditional probability: $\model(\previousplausiblepatch_{i}|\promptinitial, \previousplausiblename, \altinstruction)$

In the end, we obtain a list of plausible patches which can be given to developers for manual inspection. Different from prior \apr tools which only operate on the original buggy code to produce patches, \tech leverages additional useful information within each plausible patch to obtain more plausible patches. A plausible patch often contains useful ingredients/patterns that allowed it to pass the original test suite; therefore, instead of starting from scratch (i.e. fixing the bug again), by building on top of existing plausible patches, \chatgpt through its powerful ability to understand instructions can obtain additional plausible patches to increase the likelihood that our final list of patches contains a correct patch that fixes the bug.

\section{Experimental Design}

We evaluate \tech on the following research questions:

\begin{itemize}[noitemsep, leftmargin=*, topsep=0pt]
    \item \textbf{RQ1:} How does the performance of \tech compare against the state-of-the-art techniques for \apr?
    \item \textbf{RQ2:} How does \tech perform when used in different repair scenarios? 
    \item \textbf{RQ3:} What are the contributions of different components of \tech in improving repair effectiveness?
\end{itemize}

We first demonstrate the performance of \tech by comparing against the state-of-the-art \apr tools on the popular \dfj~\cite{just2014dfj} and \quixbugs~\cite{lin2017quixbug} repair dataset. 
Following, we closely examine each of our repair scenarios (single-line, single-hunk and single-function) with similarly evaluated baseline tools and also evaluate how our plausible patch generation step helps to improve the number of correct fixes. 
Lastly, we conduct a comprehensive ablation study on the different configurations of \tech. In particular, we look at not only the conversational aspect but also how to provide feedback along with the effect on repair performance as we change the maximum length of conversation.

\subsection{Implementation}
\label{sec:implementation}

\parabf{Repair scenarios.} In \tech, we study 3 different repair scenarios used in prior work~\cite{xia2023repairstudy}: \textbf{single-line}--fixed by replacing/adding a single line, \textbf{single-hunk}--fixed by replacing/adding a continuous code hunk and \textbf{single-function}--fixed by generating a new function to replace the original buggy version. Our initial prompts differ slightly based on the repair scenario and we provide examples of all three in Figure~\ref{fig:overview}. Note that single-hunk repair setting is studied extensively by prior learning-based \apr tools~\cite{xia2022alpharepair, ye2022rewardrepair, ye2022selfapr}.%

\parabf{Implementation.} We implement the main logic of \tech in Python by accessing the \chatgpt API endpoint~\cite{chatgptendpoint}. We use the \CodeIn{gpt-3.5-turbo-0301} model of the \chatgpt family which is the current latest model available to us. For each chosen prompt, the authors follow the best-practice guide~\cite{openaibestpractice} and manually examined a few alternative approaches with selected bugs via the Web-version of ChatGPT~\cite{chatgptweb}. We use a sampling temperature of 1 in order to get a diverse set of potential patches. Our default setting for the maximum number of repair attempts allowed (including both initial repair and plausible patch generation steps) is 200 for single-line and single-hunk \apr, and 100 for the single-function scenario.%
We use 1 few-shot example and a maximum conversation length of 3. We evaluate all generated patches on an 8-core workstation with Intel i7 10700KF Comet Lake CPU @3.80GHz and 64GB RAM, running Ubuntu 20.04.3 LTS and OpenJDK Java 64-Bit Server version 1.8.0\_312. Following prior \apr work~\cite{xia2022alpharepair, zhu2021recoder, li2020dlfix}, we use a default end-to-end timeout of 5-hours to fix one bug. In reality, our cost is far less than 5-hours due to the low number of patches sampled (<500) per bug. %

\subsection{Subject Systems}

For evaluation, we use the widely studied repair benchmark of \dfj~\cite{just2014dfj} and \quixbugs~\cite{lin2017quixbug}. \dfj is a Java benchmark collected from bug and corresponding fixes of open-source projects. Similar to prior APR tools~\cite{xia2022alpharepair, ye2022selfapr, xia2023repairstudy, ghanbari2019prapr}, we separate \dfj into 1.2 and 2.0. \dfj 1.2 consissts of 391 bugs (after removing 4 depreciated bugs) in 6 different Java projects. In this work, we following prior study~\cite{xia2023repairstudy} and categorize \dfj 1.2 into single-function (255 bugs), single-hunk (154 bugs) and single-line (80 bugs). Note that single-hunk is a subset of single-function and single-line is a subset of single-hunk bugs. We then apply our 3 proposed repair scenarios corresponding to the each of the 3 datasets. Note in RQ1, similar to prior work~\cite{xia2023repairstudy} we report the total number of bugs fixed when combining all three repair scenarios together and study each repair scenario separately in later RQs. \dfj 2.0 consists of 438 new bugs across 9 additional projects. We select only the 82 single-line bugs within \dfj 2.0 which is the main setting used in prior \apr tools for ease of comparison~\cite{xia2022alpharepair}. Furthermore, we also evaluate on the \quixbugs~\cite{lin2017quixbug} dataset which is made up of 40 buggy and fixed versions of classic programming problems in both Python and Java. All 40 bugs in \quixbugs-Python are single-function, single-hunk and single-line bugs while 40, 37, and 36 bugs in \quixbugs-Java are single-function, single-hunk and single-line bugs respectively. 

\subsection{Compared Techniques}

\parabf{Baseline techniques.} We compare \tech against state-of-the-art traditional, \nmt learning-based and \llm-based \apr baselines. We select 8 recent learning-based and \llm-based \apr baselines: \selfapr~\cite{ye2022selfapr}, \alpharepair~\cite{xia2022alpharepair}, \rewardrepair~\cite{ye2022rewardrepair}, \recoder~\cite{zhu2021recoder}, \cure~\cite{jiang2021cure}, \coconut~\cite{lutellier2020coconut}, \dlfix~\cite{li2020dlfix} and \sequencer~\cite{chen2018sequencer}. In particular, \alpharepair is a state-of-the-art \llm-based repair tool by applying pre-trained \codebert model~\cite{codex} with cloze-style \apr. Furthermore, we also include a \llm-based \apr tool built using the \codex model~\cite{feng2020codebert} (we refer to as \codexrepair) in a recent study where researchers directly applied \llm{s} for \apr without any fine-tuning~\cite{xia2023repairstudy}. \codexrepair is also studied on three repair settings used in our work which allows for more direct comparison. For traditional \apr tools, we compare against 12 selected representative techniques: \tbar~\cite{liu2019tbar}, \prapr~\cite{ghanbari2019prapr}, \avatar~\cite{liu2019avatar}, \simfix~\cite{jiang2018simfix}, \fixminer~\cite{koyuncu2020fixminder}, \capgen~\cite{wen2018capgen}, \jaid~\cite{chen2017jaid}, \sketchfix~\cite{hua2018sketchfix}, \nopol~\cite{demacro2014nopol}, \jgenprog~\cite{martinez2015automatic}, \jmutrepair~\cite{martinez2016astor}, and \jkali~\cite{martinez2016astor}. Altogether, we compare against 21 prior \apr tools. Moreover, we also evaluate against a baseline of directly sampling using the \chatgpt model to perform repair without any conversation or feedback information. We refer to this baseline as \baserepair. Since our 3 repair scenarios rely on knowing the location of the bug, we use the perfect fault localization (where the groundtruth location of the bug is given) results from prior tools. This is the preferred evaluation setting as it eliminates any differences in performance caused by running fault localization tools~\cite{ye2022rewardrepair, ye2022selfapr, wong2016fl, xia2022alpharepair, zhu2021recoder, xia2023repairstudy}. Following convention in \apr work~\cite{xia2022alpharepair, zhu2021recoder}, we directly report the fix results obtained in prior studies~\cite{xia2023repairstudy, ghanbari2019prapr, ye2022selfapr}

\parabf{Metrics.} For evaluating \tech, we use the standard metrics of \emph{plausible patches} -- passing the entire test suite and \emph{correct patches} -- semantically or syntactically equivalent to the reference developer patch. We follow common practice in \apr and manually determine the semantic equivalency to compute correct patches. Additionally, we use metric of \emph{tries} which indicates the number of samples used to obtain either a plausible or correct patch when querying \chatgpt. A lower number of tries is desirable as it reduces the time it takes to fix a bug. Finally, we also compute the \emph{dollar cost} of fixing a bug. At the time of writing, \chatgpt costs \$0.002 per every 1000 tokens~\cite{chatgptendpoint} processed or generated. Different from the number of tries, the cost can vary depending on both the number of times we query \chatgpt but also the size of input in each query. 

\section{Evaluation}

\subsection{RQ1: State-of-the-art Comparison}

\begin{table*}[!htp]\centering
\caption{Correct fixes on \dfj}\label{tab:eval_result}
\scalebox{0.85}{
\begin{tabular}{lrrrrrrrrrrr}\toprule
\textbf{Dataset} &\textbf{\makecell{\tech}} &\textbf{\makecell{BaseChatGPT}} &\textbf{\makecell{CodexRepair}} &\textbf{\makecell{AlphaRepair}} &\textbf{\makecell{SelfAPR}} &\textbf{\makecell{RewardRepair}} &\textbf{Recoder} &\textbf{TBar} &\textbf{CURE} &\textbf{\coconut} \\\midrule
Chart &15 &9 &9 &9 &7 &5 &10 &11 &10 &7 \\
Closure &37 &23 &30 &23 &19 &15 &21 &16 &14 &9 \\
Lang &21 &15 &22 &13 &10 &7 &11 &13 &9 &7 \\
Math &32 &25 &29 &21 &22 &19 &18 &22 &19 &16 \\
Mockito &6 &6 &6 &5 &3 &3 &2 &3 &4 &4 \\
Time &3 &2 &3 &3 &3 &1 &3 &3 &1 &1 \\
\midrule
\textbf{D4J 1.2} &\textbf{114} &80 &99 &74 &64 &50 &65 &68 &57 &44 \\
\midrule
\textbf{D4J 2.0} &\textbf{48} &25 &31 &36 &31 &25 &11 &8 &- &- \\
\bottomrule
\end{tabular}
}
\end{table*}

\begin{table}[!htp]\centering
\caption{Correct fixes on \quixbugs}\label{tab:eval_result_quixbugs}
\scalebox{0.85}{
\begin{tabular}{lrrrrrr}\toprule
\multirow{2}{*}{\textbf{\quixbugs}} &\multirow{2}{*}{\textsc{\textbf{\makecell{Chart\\Repair} }}} &\multirow{2}{*}{\textbf{\makecell{Base\\ChatGPT}}} &\multirow{2}{*}{\textbf{\makecell{Codex\\Repair}}} &\multirow{2}{*}{\textbf{\makecell{Alpha\\Repair}}} &\multirow{2}{*}{\textbf{\coconut}} \\
& & & & & \\\midrule
\textbf{Python} &\textbf{40} &40 &40 &27 &19 \\
\textbf{Java} &\textbf{40} &40 &38 &28 &13 \\
\bottomrule
\end{tabular}
}
\end{table}

\begin{table}[!htp]\centering
\caption{Correct fixes using three repair settings}\label{tab:repair_setting}
\scalebox{0.85}{
\begin{tabular}{l|rrr|rrr|rrrr}\toprule
\multirow{2}{*}{\textbf{Tools}} &\multicolumn{3}{c}{\textbf{D4J 1.2}} &\multicolumn{3}{c}{\textbf{Quixbugs-Py}} &\multicolumn{3}{c}{\textbf{Quixbugs-J}} \\\cmidrule{2-10}
&\textbf{SL} &\textbf{SH} &\textbf{SF} &\textbf{SL} &\textbf{SH} &\textbf{SF} &\textbf{SL} &\textbf{SH} &\textbf{SF} \\\midrule
\textbf{\tech} &\textbf{57} &\textbf{79} &\textbf{76} &\textbf{39} &\textbf{40} &\textbf{40} &\textbf{36} &\textbf{37} &\textbf{39} \\
\textbf{\baserepair} &41 &55 &45 &38 &37 &35 &33 &36 &39 \\
\textbf{\codexrepair} &39 &62 &63 &39 &39 &37 &34 &34 &32 \\
\bottomrule
\end{tabular}
}
\end{table}

We first compare \tech against the state-of-the-art \apr tools. Table~\ref{tab:eval_result} shows the number of bugs fixed on \dfj 1.2 and 2.0 by the top baseline tools as well as \baserepair -- only using \chatgpt without any test failure information and conversation. We first observe that \tech can improve over the baseline of just using the \chatgpt model with 34 and 23 more bug fixes on \dfj 1.2 and 2.0 respectively. This improvement is obtained by successfully leveraging the conversational aspect of \chatgpt model to provide immediate feedback using both previous incorrect or plausible patches and test failure information. Interestingly, we also observe that prior tools such as \alpharepair which uses a much smaller \llm (\codebert) can perform better than \codexrepair in cases like single-line repair on \dfj2.0 due to the use of repair-specific templates compared with pure code infilling. In fact, \tech demonstrates for the first time that \llm-based \apr without any repair templates can achieve top performance on \dfj. In total, \emph{\tech is able to achieve 114 and 48 correct bug fixes on \dfj 1.2 and 2.0 respectively, with 15 and 17 more than the current state-of-the-art \apr tool.} Calculating the total cost of query \chatgpt, we can \emph{fix 162 out of 337 bugs for \$0.42 each!}
While the 114 and 48 fixes are achieved by combining three repair settings together, \tech still generates far less patches (<500 in total per bug) compared to prior learning-based tools which can generate up to 10,000 patches per bug~\cite{lutellier2020coconut, jiang2021cure}. Similarly in Table~\ref{tab:eval_result_quixbugs}, \tech is able to correctly fix all bugs within the  \quixbugs-Java and -Python datasets, beating out all top-performing techniques.

\begin{figure}[t]
    \captionsetup{justification=centering}
    \centering
    \includegraphics[keepaspectratio=true,width=0.5\linewidth]{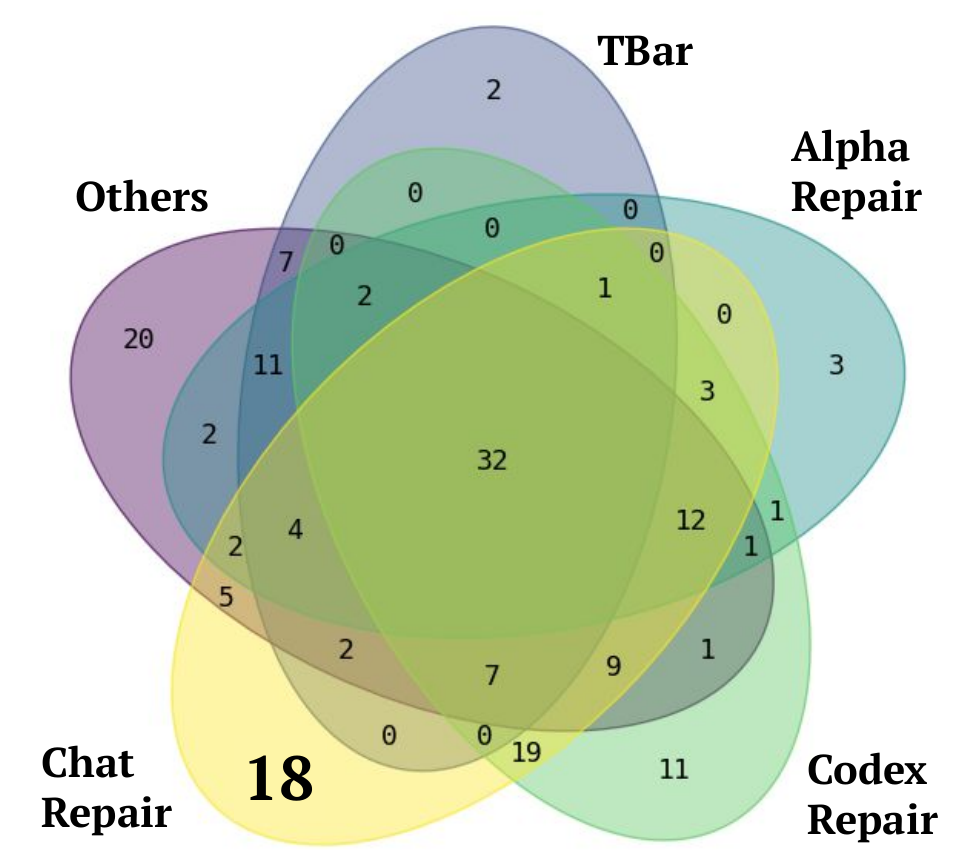}
    \caption{Bug fix Venn diagram on \dfj 1.2}
    \label{fig:venn}
\end{figure}

\begin{figure}[t]
    \captionsetup{justification=centering}
    \centering
    \includegraphics[keepaspectratio=true,width=0.8\linewidth]{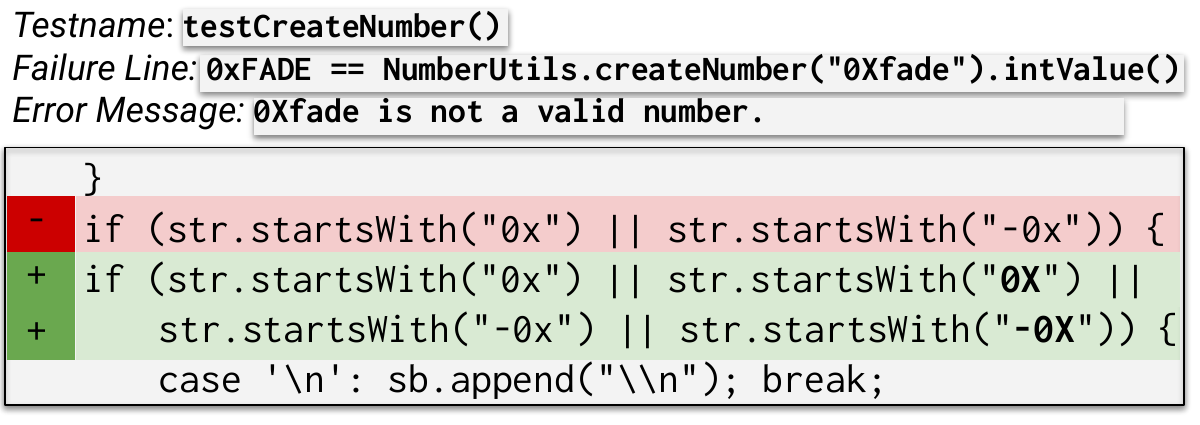}
    \caption{Unique bug fixed in \dfj 1.2}
    \label{fig:dfj12_bug_fix}
\end{figure}

\begin{figure}[t]
    \captionsetup{justification=centering}
    \centering
\includegraphics[keepaspectratio=true,width=0.8\linewidth]{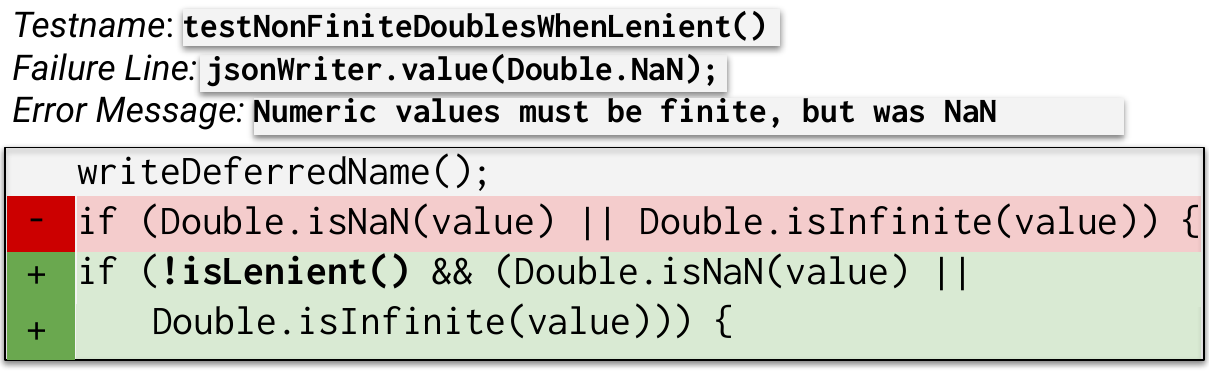}
    \caption{Unique bug fixed in \dfj 2.0}
    \label{fig:dfj2_bug_fix}
\end{figure}

Figure~\ref{fig:venn} shows the Venn diagram of the bug fixed by all studied baselines and \tech on \dfj 1.2. We select the 3 top baselines in terms of the number of bugs fixed and group all other studied \apr tools (not just the top-performing ones in Table~\ref{tab:eval_result}) as ``Other''. We see that \emph{\tech can provide the correct patch for 18 unique bugs that no prior approach is able to fix so far on \dfj 1.2}. To illustrate the power of \tech, we show an example bug (\CodeIn{Lang-16}) in \dfj 1.2 that is only fixed by \tech in Figure~\ref{fig:dfj12_bug_fix}. The fix is to append two additional conditions of starting with either \CodeIn{"-0X"} or \CodeIn{"0X"} referring to hexadecimal representation of a number. This bug is difficult to fix, since the strings are not commonly found in either bug-fix training data (\nmt-based) or in pre-training data (\llm-based). In order to generate these condition, the \apr tool needs to understand the expected behavior and what other usage inputs may look like. In fact, one of the condition (\CodeIn{"0X"}) is directly used in the failing test where the test tries to create a number of \CodeIn{"0Xfade"}. \tech is able to leverage this relevant test code information and generate the string used in the test as a condition. Furthermore, the new negative variant \CodeIn{"-0X"} can also be easily generated by \tech as \chatgpt is able to learn from the original buggy line which also contains pairs of negative and positive conditions. Combining both conditions together, \tech is able to obtain the correct patch that fixes this bug. 

Another bug (\CodeIn{Gson-15}) that can only be fixed by \tech is presented in Figure~\ref{fig:dfj2_bug_fix} from \dfj 2.0. The fix requires another unique condition of \CodeIn{!isLenient()}. To make things worse, this usage of the function is not found within the original buggy function context. As such, it can be extremely difficult for prior learning-based \apr tools to fix since there are no example usages of the condition within the context. However, we observe that the failing test is named \CodeIn{testNonFiniteDoublesWhenLenient} where the word lenient directly appears. \chatgpt, through looking at the failing test name, can understand the semantic meaning of the test which in this case is to test a particular setting with \CodeIn{lenient = true} and generate the correct fix line to check for this unique setting. This example further shows the power of \tech in leveraging previously ignored semantic information within failing tests to directly guide the repair process. 

\subsection{RQ2: Repair Scenarios} 

Next, we take a look at each of our three repair settings (single-line, single-hunk and single-function) in more detailed. For this section, we focus our analysis against \baserepair using \chatgpt without any test failure information or conversation, and \codexrepair which is the best performing \llm-based \apr and has also been evaluated on the three repair settings that we use. 

Table~\ref{tab:repair_setting} shows the results of \tech against the two baselines on \dfj 1.2 and two \quixbugs datasets. Interestingly, we first observe that the base \chatgpt model performs even slightly worse than \codexrepair on the real-world benchmark of \dfj 1.2. We theorize that this is because \chatgpt is not designed or directly fine-tuned for code generation like \codex. As such, directly using \chatgpt in a similar fashion to prior \llm-based \apr tools that solely sample from the same initial prompt without additional information does not yield impressive improvements~\cite{ye2023comprehensive}. On the other hand, by using \tech, which combines the powerful dialogue/instruction understanding ability of \chatgpt with dynamic feedback, \tech is able to better leverage the previously ignored test failure information and earlier patch attempts to better perform \apr. \codex on the contrary, is designed mainly for code completion and lacks the ability to be used in a conversational manner. In summary, for each individual repair setting, \tech is able to achieve the highest number of bugs fixed compared to both state-of-the-art \codexrepair and running base \chatgpt.

\begin{figure}[t]
    \captionsetup{justification=centering}
    \centering
    \includegraphics[keepaspectratio=true,width=0.8\linewidth]{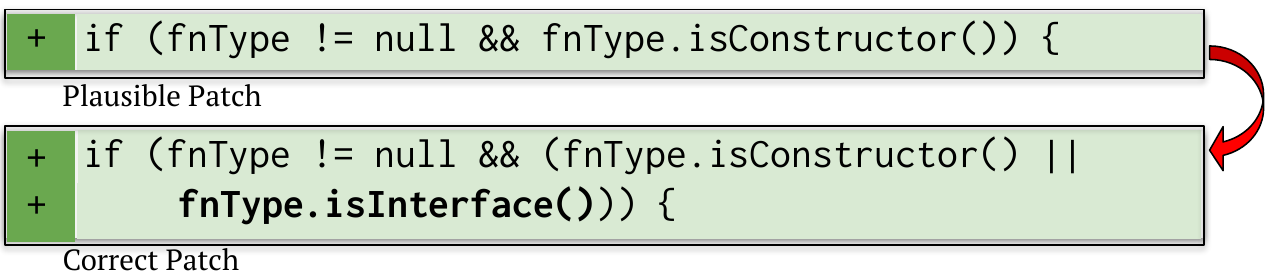}
    \caption{Plausible generation example}
    \label{fig:ptc_example}
\end{figure}

Additionally, the improvement in more correct fixes does not only come from the conversational and validation feedback aspect but is also contributed by our plausible patch generation step. Recall that once a plausible patch is generated, we directly use that patch to generate more plausible patches by asking \chatgpt to provide other variations of the patch. In summary, plausible patch generation is able to add on average an additional 9.4, 16.6, 5.5 plausible patches, and improve the number of correctly fixed bugs in single-line, single-hunk, and single-function repair scenarios by 4, 7, 2 respectively on \dfj 1.2. This improvement demonstrates the usefulness of our proposed approach in leveraging the important information in plausible patches to generate more patches leading to a correct fix.  

Figure~\ref{fig:ptc_example} shows an example of a correct fix (\CodeIn{Closure-125}) by \tech which was initially only plausible and then became correct after guiding \chatgpt to learn from the earlier plausible patch. We see that the initial plausible patch produced by \tech is indeed able to pass the developer tests by checking if \CodeIn{fnType} is a constructor. However, the testsuite does not cover all corner cases and the actual correct fix involves checking an additional condition of an interface. By using the plausible patch generation, \tech does not have to start from scratch (using only the buggy code) but instead can build on the knowledge already obtained in the first plausible patch. In this bug fix, \tech adds the additional condition required to correctly fix by learning from the original plausible patch.

\subsection{RQ3: Configurations of \tech}

We investigate the different configurations of \tech. Specifically we examine the important parameters of (1) initial prompt used, (2) feedback response provided and (3) maximum conversation length. Due to the substantial cost of invoking the \chatgpt API multiple times for each dimension of our ablation study, we focus on the 80 single-line bugs within \dfj 1.2. Also, we analyze the number of plausible fixes produced instead of correct fixes in this RQ due to the intensive manual efforts involved in patch inspection. Each of our ablation experiments uses the default setting described in Section~\ref{sec:implementation} except we use zero-shot (not providing any prior bug fix examples) by default since it can best illustrate the effect of individual components and make it easier for studying the impact of few-shot examples.

\subsubsection{Initial Prompt}

\newcommand{\basemessage}{BasePrompt\xspace}
\newcommand{\testnamerror}{TestName+ErrMsg\xspace}
\newcommand{\testfunction}{TestName+ErrMsg+TestBody\xspace}
\newcommand{\testnamerrorline}{TestName+ErrMsg+FailLine\xspace}
\newcommand{\basesystem}{\CodeIn{You are a helpful assistant}\xspace}
\newcommand{\aprsystem}{\CodeIn{You are an APR tool}\xspace}

\begin{table}[!htp]\centering
\caption{Initial prompt variations}
\label{tab:initial_prompt_ab}
\scalebox{0.85}{
\begin{tabular}{lrrrr}\toprule
\textbf{Initial Prompt} &\textbf{\#P} &\textbf{Avg. \# tries} &\textbf{Avg. \$} \\\midrule
\basemessage &55 &22.53 &\$0.069 \\
\testnamerror &59 &22.47 &\$0.072 \\
\testnamerrorline &\textbf{64} &\textbf{21.86} &\textbf{\$0.061} \\
\testfunction &61 &23.42 &\$0.083 \\
\midrule
\basesystem &\textbf{64} &24.17 &\$0.074 \\
\aprsystem &\textbf{64} &\textbf{21.86} &\textbf{\$0.061} \\
\midrule
0-shot &64 &21.86 &\textbf{\$0.061} \\
1-shot &\textbf{65} &9.91 &\$0.072 \\
2-shot &\textbf{65} &\textbf{9.87} &\$0.085 \\
\bottomrule
\end{tabular}
}
\end{table}

In addition to our default initial prompt given to \chatgpt, we also evaluate several alternative variations. Each variation attempts to illustrate some key aspects of information which can be helpful for \chatgpt during the repair process. Table~\ref{tab:initial_prompt_ab} shows the results of the different initial prompts. Row \textbf{\basemessage} refers to the prompt where we only indicate the code contains a bug and asks the model to provide a fix, \textbf{\testnamerror} includes both the failing test name (e.g., \CodeIn{testGetCategoryIndex}) and test failure error message (e.g., \CodeIn{NullPointerException}), \textbf{\testnamerrorline} additionally includes the exact line where the failure occurred within the test (e.g., \CodeIn{assertEquals(-1, empty.getCategoryIndex("ABC"));}) and \textbf{\testfunction} additionally uses the entire failing test function body instead of just the failure code line. 

First, we observe that the base initial prompt of only providing with the buggy code and asking it to generate a patch performs the worst in terms of the number of bugs fixed. We see that by adding auxiliary information such as failing test name and error message, we can further improve the repair performance. Tests that are well named can provide semantic meaning of the test. Error messages also offer unique insights regarding the nature of the test failure (e.g., null point exception, array out of bound checks) and can directly motivate a potential correct patch. Furthermore, we remark that providing the exact line within the test where the failure occurred can also improve repair performance. Such lines may include assertions showing desired results (e.g., numerical comparison) or statements that triggered the exceptions or crashes (e.g., field dereferences). This additionally gives concrete hints to \chatgpt on how to fix the specific bug. Moreover, we observe that the prompt which includes the entire test function code also performs well in terms of the number of bugs fixed. However, we see that on average it costs the most compared to the other initial prompts used. This is because the model incurs additional cost to process the entire test function code for each repair attempt, which can be largely depending on the size of the test function and may contain test code irrelevant to the bug. As such, a more concise prompt which includes just the failing test code line can already achieve effective repair performance while being economic.

Table~\ref{tab:initial_prompt_ab} also shows several other parameters of the initial input apart from the included test failure information. Row \textbf{\basesystem} uses the default system message (Section~\ref{sec:approach}) by \chatgpt and \textbf{\aprsystem} (we use the full name of \apr) is our modified system message. While the number of fixes is similar between the two system messages, we observe that by aligning the system message with the task we want to solve -- program repair, the model can arrive at the plausible patch faster (less tries) since it can faster understand the task it is trying to solve. As such we can reduce the cost of \tech by designing specific system messages. Furthermore, we evaluate the effect of having few-shot examples of bug fixes before the target buggy code input in Table~\ref{tab:initial_prompt_ab}. We observe that by providing \chatgpt with some examples of prior bug fixes, we obtain a slight increase in number of plausible patches while at the same time drastically reducing the number of tries used to fix the bug. Few-shot examples, similar to the system message, can get the model familiar for bug fixing by understanding the task and input/output formats.

\subsubsection{Feedback Response}

\newcommand{\basefeedback}{BaseFeedback\xspace}
\newcommand{\dynamic}{Dynamic\xspace}

\begin{table}[!htp]\centering
\caption{Feedback response variations}
\label{tab:feedback_response_ab}
\scalebox{0.85}{
\begin{tabular}{lrrrr}\toprule
\textbf{Feedback Response} &\textbf{\#P} &\textbf{Avg. \# tries} &\textbf{Avg. \$} \\\midrule
\basefeedback &58 &23.12 &\$0.071 \\
\testnamerror &61 &22.48 &\$0.073 \\
\testnamerrorline &62 &24.71 &\$0.074 \\
\dynamic &\textbf{64} &\textbf{21.86} &\textbf{\$0.061} \\
\bottomrule
\end{tabular}
}
\end{table}

Another important aspect of \tech's design is the feedback response we provide to the model. Similar to the initial prompt design, we also consider multiple different ways we can provide feedback to the model. Table~\ref{tab:feedback_response_ab} shows the results of the different feedback response variants. Row \textbf{\basefeedback} means we only tell the model that the generated patch is not correct without any additional feedback. Similar to the initial prompt construction, \textbf{\testnamerror} includes both the failing test name and test failure error message, \textbf{\testnamerrorline} additionally includes the exact line where the failure occurred within the test. Different from initial prompt construction, \textbf{\dynamic} is our default approach where we only provide the test name/error/line if the new generated patch has a different failure than the original (Section~\ref{sec:conversation}). This allows us to more concretely inform \chatgpt if it has made some progress in fixing a bug (e.g., patch no longer crashes with null-pointer exception but fails on some other test).

Initially, we see that the base response message achieves the worst result in number of bugs fixed. Similar to the behavior of the initial prompts, we can improve performance by adding the name of the failing test, error message along with the exact failure line from the failing test. Additionally, we can further improve performance by implementing the dynamic feedback response. Since in the initial prompt we already provide \chatgpt with the failing test name, error message, and failing line, in dynamic feedback response, we only provide new data if the generated patch contains a different failing information. This allow us to make more use of the conversational aspect by referring to a previous message. Furthermore, it can reduce the cost as we only produce a short concise message if the patch does not make additional progress. 

\subsubsection{Conversation Length}
\begin{figure}[t]
    \captionsetup{justification=centering}
    \centering
    \includegraphics[keepaspectratio=true,width=0.85\linewidth]{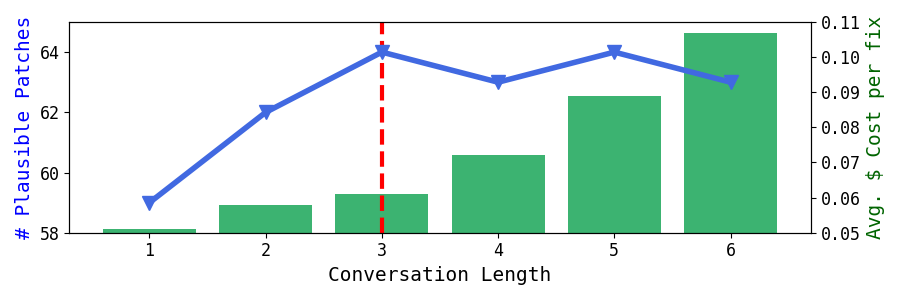}
    \caption{Effect of maximum conversation length}
    \label{fig:length_ab}
\end{figure}

Figure~\ref{fig:length_ab} shows performance in both number of plausible fixes and the average dollar cost to fix a bug across different maximum conversation length. Recall from Section~\ref{sec:conversation} that the maximum conversation length dictates the amount of history/feedback within each individual repair conversation, where length = 1 is equivalent to sampling using the initial prompt without any feedback. We observe that by directly sampling from the \chatgpt without any conversation, we achieve the lowest number of plausible fixes. As we add the conversation/feedback element of \tech, we see that we can improve the number of plausible patches. Compared with sampling from the same prompt over and over again, by using \tech in a conversational manner, the model can learn from its previous mistakes along with the concise test failure feedback information to generate more plausible patches. We also notice that the model can retain its performances as we increase the conversation length to be higher (i.e., 5 and 6). However, we see that compared with a lower conversation length (3), the higher conversation lengths incurs a much higher cost in fixing a bug. The reason is that as we increase the length, the amount of history/context (tokens) processed by the model will be higher, leading to higher cost per bug fixed. Our default conversation length of 3 serves as a good balance between cost and the number of bugs fixed.

\section{Threats to Validity}

\parabf{Internal.} The first internal threat comes from the manual validation used to determine the correctness of the plausible patches compared with the reference developer patch. To address this, following prior work~\cite{xia2022alpharepair, xia2023repairstudy, ye2022selfapr, ye2022rewardrepair, zhu2021recoder, jiang2021cure}, we carefully examined and discussed each patch.

Another threat to validity comes from the data leakage of reference developer patches being part of the original training data of \chatgpt. Since \chatgpt is a proprietary model and can only be accessed through API, we do not have access to the exact training data used. To address this, we follow prior work~\cite{xia2022alpharepair} and first compute the number of correct patches generated by \tech which was the same as the reference developer patch on \dfj 1.2. We found that out of 212 (adding all correct patches from our three repair scenarios) correct patches, 77 of them is the same as reference developer fix (36\%). In addition, even if we remove all correct patches (77) which are the same as the reference developer patch, \tech is still able to generate the correct patch for 12 unique bugs that none of the prior approaches can fix. Furthermore, compared to the base \chatgpt repair baseline which uses the same underlying model, \tech is able to drastically improve its performance (34 more correct fixes) showing that the result gained by \tech is not simply due to memorizing the training data. To completely address this threat, we would need to retrain \chatgpt from scratch which would be infeasible for an academic project. 

\parabf{External.} The main external threat to validity comes for our evaluation datasets used. The improvement obtained by \tech may not generalize to other repair datasets. To address this, we evaluate not only on the popular \dfj 1.2 dataset but also on \dfj 2.0 and two \quixbugs datasets to demonstrate the generalizability.

\section{Conclusion}

We propose \tech{} -- the first fully automated conversation-driven \apr tool which leverages the newly developed \chatgpt model to perform repair. \tech learns from both previously incorrect and plausible patches and utilizes test failure information to provide immediate and dynamic feedback to the model to generate a new patch. Through our conversational repair paradigm, \tech is able to achieve the new state-of-the-art performance of 114 and 48 bugs (15 and 17 more than best-performing baseline) on \dfj 1.2 and 2.0 respectively.

\bibliographystyle{ACM-Reference-Format}
\bibliography{references}

\end{document}